\documentclass{article}
\usepackage{spconf,amsmath,graphicx,xcolor}
\usepackage[ruled, lined, linesnumbered, commentsnumbered, longend]{algorithm2e}
\usepackage{hyperref}

\title{Speech Emotion Recognition with ASR Transcripts: A Comprehensive Study on Word Error Rate and Fusion Techniques}
\name{Yuanchao Li, Peter Bell, Catherine Lai}
\address{Centre for Speech Technology Research, University of Edinburgh}

\begin{document}
%
\maketitle
\begin{abstract}
Text data is commonly utilized as a primary input to enhance Speech Emotion Recognition (SER) performance and reliability. However, the reliance on human-transcribed text in most studies impedes the development of practical SER systems, creating a gap between in-lab research and real-world scenarios where Automatic Speech Recognition (ASR) serves as the text source. Hence, this study benchmarks SER performance using ASR transcripts with varying Word Error Rates (WERs) from eleven models on three well-known corpora: IEMOCAP, CMU-MOSI, and MSP-Podcast. Our evaluation includes both text-only and bimodal SER with six fusion techniques, aiming for a comprehensive analysis that uncovers novel findings and challenges faced by current SER research. Additionally, we propose a unified ASR error-robust framework integrating ASR error correction and modality-gated fusion, achieving lower WER and higher SER results compared to the best-performing ASR transcript. These findings provide insights into SER with ASR assistance, especially for real-world applications.
\end{abstract}
\section{Introduction}
\label{sec:intro}

Emotion in speech is conveyed through both what is said (i.e., text) and how it is said (i.e., audio). Recent studies on Speech Emotion Recognition (SER) consistently highlight the benefits of combining textual and acoustic features \cite{ghosh2022mmer,sebastian2019fusion, yoon2018multimodal,priyasad2023dual}. Despite the advancements in SER over the years, its practical application in everyday contexts remains limited. One significant challenge is the reliance on human-annotated text, known as gold-standard manual transcripts. In contrast, transcripts generated by Automatic Speech Recognition (ASR) systems, even state-of-the-art ones, often suffer from high Word Error Rates (WERs), particularly in emotional speech \cite{li23ea_interspeech}. As a result, findings from lab settings may not effectively translate to real-world scenarios, impeding the true progress of SER.

To address this gap, it is crucial to investigate the impact of imperfect text data generated by ASR on SER. While previous studies have explored the question of \textit{``how ASR performance affects SER''}, the findings have been inconsistent and not comprehensive. For example, Schuller et al. observed that a WER of over 30\% resulted in an SER accuracy drop of less than 3\% on the FAU Aibo Emotion Corpus \cite{batliner2008releasing}, whereas Li et al. reported a nearly 10\% accuracy drop with a 30\% WER on IEMOCAP \cite{li23ea_interspeech}. Furthermore, while multimodal fusion techniques have been proposed and extensively studied to enhance emotion recognition performance, their compatibility with ASR transcripts remains underexplored \cite{wang2023crossattention,gao2021domain,yao2022modality}.

In this study, we begin with benchmarking SER performance using ASR transcripts. We generate ASR transcripts from multiple ASR systems to achieve varying WERs. Subsequently, SER is conducted exclusively on these transcripts to evaluate the impact of ASR errors on text-only SER performance. Following this, we integrate audio features and perform bimodal SER to investigate whether fusion techniques, effective with ground-truth text, also demonstrate efficacy with ASR transcripts (i.e., their resilience to ASR errors). Building on the benchmark study, we propose an ASR error-robust SER framework that integrates ASR error correction with modality-gated fusion to address both WER and fusion challenges in SER. The contributions of this work can be summarized as follows:

$\bullet$ As the first work to benchmark SER performance with ASR transcripts, we examine a wide range of models, varying WERs, and multiple fusion techniques. This sets the stage for advancing SER research with ASR integration, enhancing its practical applications.

$\bullet$ We propose an ASR error-robust framework that integrates two-stage ASR error correction and dynamic modality-gated fusion, aiming to reduce word errors and mitigate the negative effects of high WER on SER performance.

\section{Related Work}
Over the past several years, significant progress has been achieved in SER through the utilization of ASR transcripts. Specifically, in chronological order:

Yoon et al. (2018) proposed a deep dual recurrent encoder model that simultaneously utilizes audio signals and text data from the Google Cloud Speech API \cite{yoon2018multimodal}. Sahu et al. (2019) utilized two commercial ASR systems to generate transcripts for bimodal SER (audio + text), resulting in a relative loss of unweighted accuracy compared to ground-truth transcripts \cite{sahu19_interspeech}. Li et al. (2020) introduced a temporal alignment mean-max pooling mechanism to capture subtle and fine-grained emotions in utterances, alongside a cross-modality excitement module for sample-specific adjustments on embeddings \cite{li2020learning}.

Santoso et al. (2021) proposed using a confidence measure to adjust the importance weights in ASR transcripts based on the likelihood of a speech recognition error in each word, effectively mitigating the effects of ASR errors on SER performance \cite{santoso2021speech}. Wu et al. (2021) introduced a dual-branches model for ASR error robustness, with a time-synchronous branch combining speech and text modalities and a time-asynchronous branch integrating sentence text embeddings from context utterances \cite{wu2021emotion}. Shon et al.(2021) generated pseudo labels on ASR transcripts for semi-supervised speech sentiment analysis \cite{shon2021leveraging}.

Inspired by human perception mechanisms, Li et al. (2022) proposed hierarchical attention fusion of audio features, ASR hidden states, and ASR transcripts, achieving similar SER performance as ground-truth text \cite{li2022fusing}. Lin et al. (2023) explored complementary semantic information from audio to mitigate the impact of ASR errors, using an attention mechanism to calculate weighted acoustic representations fused with text representations of ASR hypotheses \cite{lin2023robust}. He et al. (2024) incorporated two auxiliary tasks, ASR error detection and ASR error correction, to enhance the semantic coherence of ASR text. They introduced a novel multi-modal fusion method to learn shared representations across modalities \cite{he2024mf}. Feng et al. (2024) fused audio with ASR transcript from a powerful ASR model and identified that ASR-generated output delivers competitive SER performance compared to ground-truth transcripts \cite{feng2024foundation}.

While these studies have highlighted the effectiveness of integrating text features from ASR transcripts, there is still a lack of understanding regarding how WER and fusion techniques impact SER. Therefore, we undertake a benchmark study utilizing diverse ASR models and fusion techniques, conducting SER on various emotion corpora to get a clearer picture of the effect of transcription errors on SER. Furthermore, drawing from the existing literature, we develop an ASR error-robust framework aimed at mitigating the adverse effect of WER\footnote{\href{https://github.com/yc-li20/SER-on-WER-and-Fusion}{https://github.com/yc-li20/SER-on-WER-and-Fusion}}.

\section{ASR Models and Emotion Corpora}
\noindent{\textbf{ASR models}}. We adopt the following 11 models as they are widely used in the speech area and can provide varying WERs. \\
\textbullet\ \textit{Wav2Vec2-base-\{100h,960h\}} \\
\textbullet\ \textit{Wav2Vec2-large-960h} \\
\textbullet\ \textit{Wav2Vec2-large-960h-lv60-self} \\
\textbullet\ \textit{HuBERT-large-ls960-ft} \\
\textbullet\ \textit{WavLM-libri-clean-100h-base-plus} \\
\textbullet\ \textit{Whisper-\{tiny, base, small, medium, large-v2\}.en}

\noindent{\textbf{Emotion corpora}}.
To ensure generalizability, we utilize three corpora: IEMOCAP \cite{busso2008iemocap}, CMU-MOSI \cite{Zadeh2016}, and MSP-Podcast \cite{lotfian2017building}, to include diverse speech conditions and various evaluation metrics for both discrete and continuous emotions.

IEMOCAP consists of five sessions of scripted and improvised dialogues conducted in a research lab. Our study focuses on four emotion classes: angry, happy (including excited), neutral, and sad. We exclude utterances with blank transcripts from this corpus, resulting in a total of 5,500 utterances.
CMU-MOSI comprises 2,199 monologue video clips sourced from YouTube, each annotated with sentiment scores ranging from -3 to 3.
MSP-Podcast contains English speech extracted from podcast recordings, annotated with valence, arousal, dominance scores in the range of 1 to 7, as well as categorical emotion labels. We use its Release 1.11 version and evaluate its performance on the Test1 set. We exclude utterances without an emotion label, resulting in a total of 104,663 utterances.

Following the literature, we compute \textbf{Acc4} (four-class accuracy) for IEMOCAP, Concordance Correlation Coefficient (\textbf{CCC}) for MSP-Podcast, and \textbf{Acc2} (binary accuracy), \textbf{Acc7} (seven-class accuracy), and Mean Absolute Error (\textbf{MAE}) for CMU-MOSI. Detailed explanations are omitted as these metrics are commonly used for their respective corpora \cite{wang2023crossattention,wu2021emotion}.

We initially considered including MELD \cite{poria2019meld} (TV sitcom data) in our investigation. However, its WERs are nearly double those of the other three corpora, ranging from 30\% to 65\%. Given that conducting SER using transcripts with such poor ASR performance is impractical in real-world scenarios, we decided to focus on the other three corpora for our subsequent study. Nevertheless, we present the WERs of MELD for reference (see Table~\ref{tab:wer_ser}).

\section{Benchmark Study}

\subsection{SER Modeling}
We employ RoBERTa-base as the text encoder. Given that transcripts generated by models other than \textit{Whisper} lack punctuation, we remove punctuation from the transcripts of \textit{Whisper} models to ensure a fair comparison. All letters are lowercased for consistency.
A backbone SER model is built for all corpora. Since our goal is not to achieve state-of-the-art performance, the model simply comprises two dense layers: the first encodes RoBERTa output of dimension 768 into hidden states of dimension 128, and the second further encodes it into a dimension of 16. We use ReLU as the activation function between the dense layers. In the case of IEMOCAP, we apply one output layer with Softmax activation for classification. For MSP-Podcast and CMU-MOSI, which involve regression tasks, no final output activation is applied. We set the learning rate as 5e-4 for IEMOCAP and CMU-MOSI, and 1e-4 for MSP-Podcast, using the AdamW optimizer. The weight decay is set as 1e-5, and the batch size is 64. For training, we employ five-fold cross-validation on IEMOCAP (100 epochs) and follow the official training/validation/testing sets on CMU-MOSI (100 epochs) and MSP-Podcast (30 epochs). The random seeds are kept consistent across all experiments.

\subsection{Benchmarking SER with WER}
Firstly, we present the WERs and corresponding SER performance based on transcripts from different ASR models. From Table~\ref{tab:wer_ser}, we observe that:

\textbf{1) SER performance generally decreases as WER increases.} On IEMOCAP and MOSI, there is nearly a 10\% accuracy decrease with WERs around 40\%, regardless of Acc2, Acc4, or Acc7. However, exceptions exist. For example, on IEMOCAP, \textit{Whisper-tiny} produces higher accuracy than its neighboring models, even though the WERs are similar. Additionally, \textit{W2V960-large-self} yields worse accuracy compared to its neighboring models, despite having a relatively lower WER compared to others. The same phenomenon can also be found with CMU-MOSI: the Acc2 of \textit{W2V100} is higher than that of \textit{WavLM plus}, and the Acc2 of \textit{Whisper-large} is lower than that of \textit{Whisper-medium}. This might be due to certain words being misrecognized as words that have little effect on or even positively contribute to their ground-truth emotion labels. Note that the values of our Acc4 are higher than those in \cite{li23ea_interspeech}, possibly because of differences in training models, language models (RoBERTa vs. BERT), and word embeddings (full hidden states vs. pooler output).

\begin{table}[ht!]
    \centering
    \caption{Benchmarking SER performance based on ASR transcripts across corpora. $\uparrow$: higher the better. $\downarrow$: lower the better. \textbf{\textcolor{red}{Red bold}}: better performance than the ground truth.}
    \includegraphics[width=0.47\textwidth]{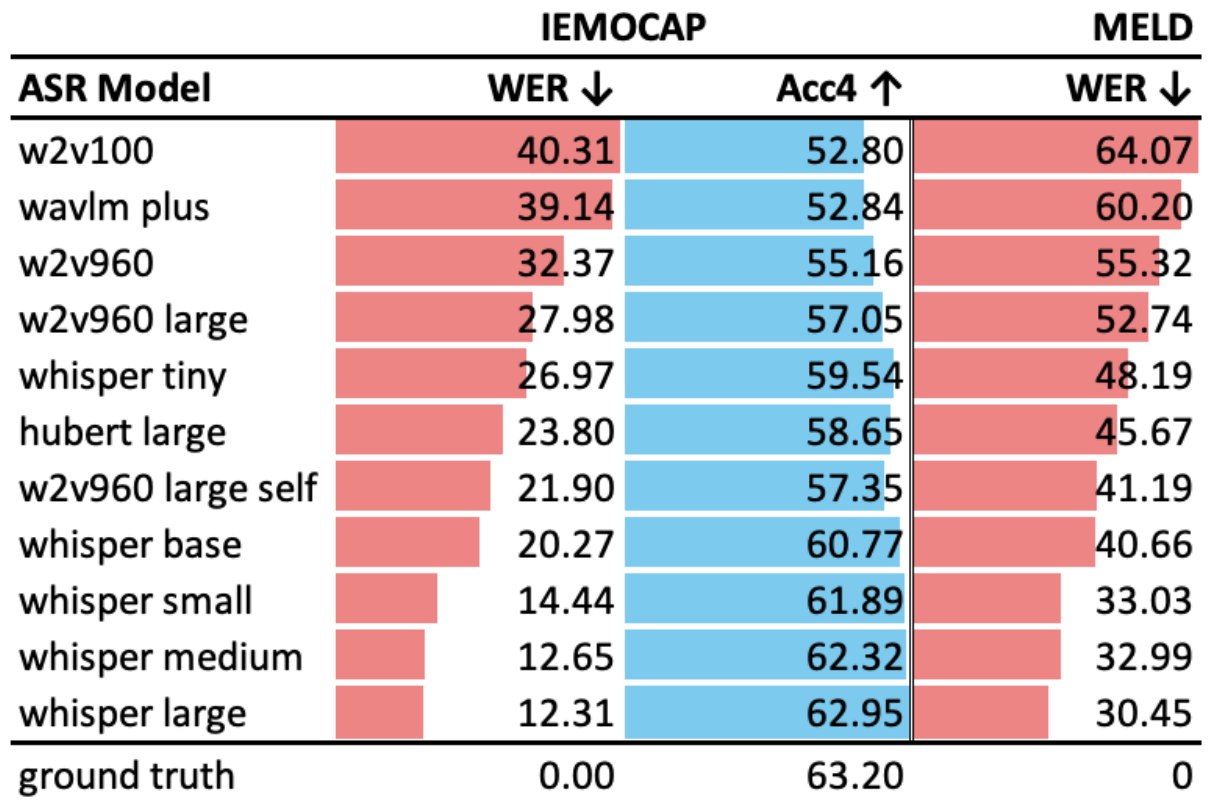}\vspace{3pt}
    \includegraphics[width=0.47\textwidth]{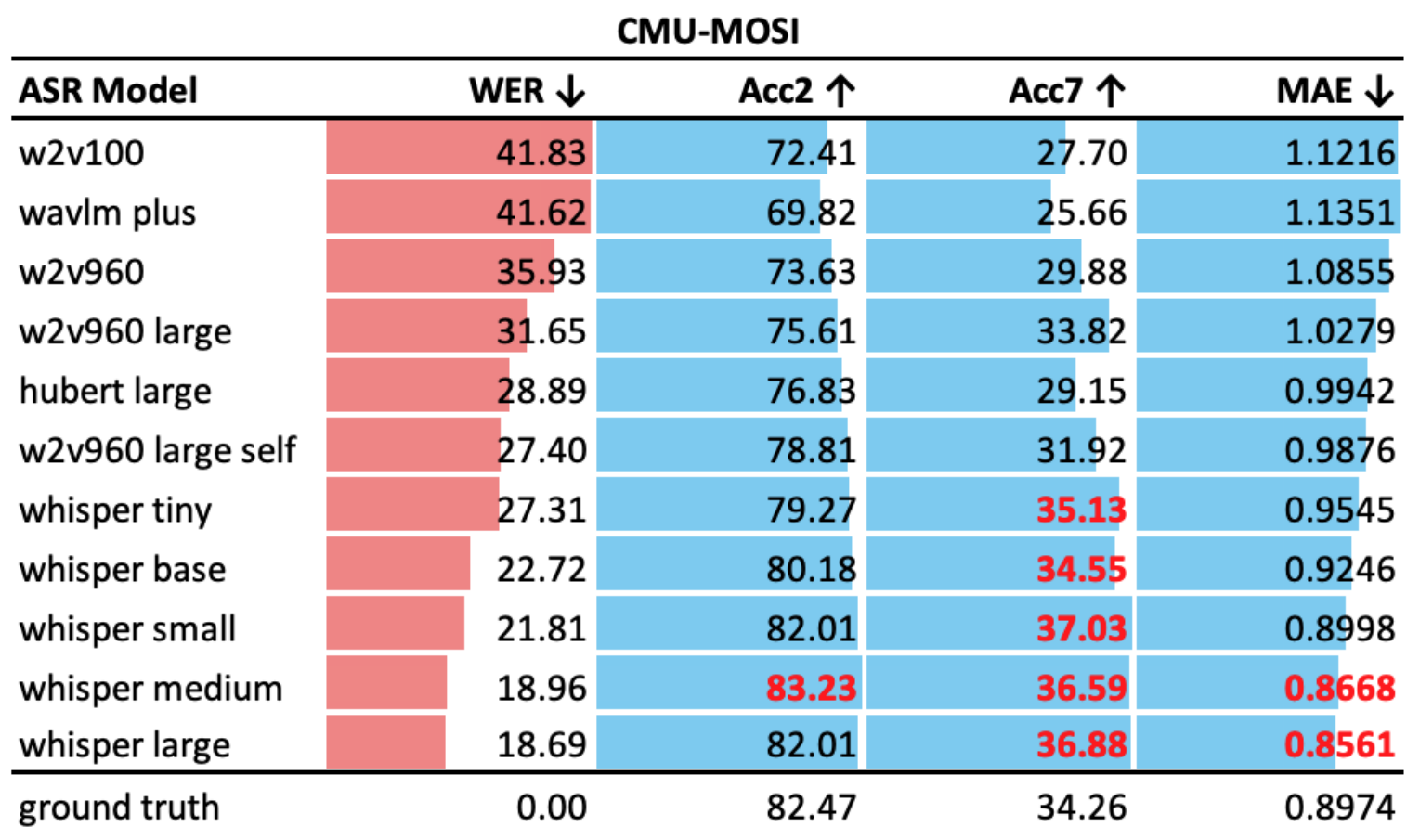}\vspace{3pt}
    \includegraphics[width=0.47\textwidth]{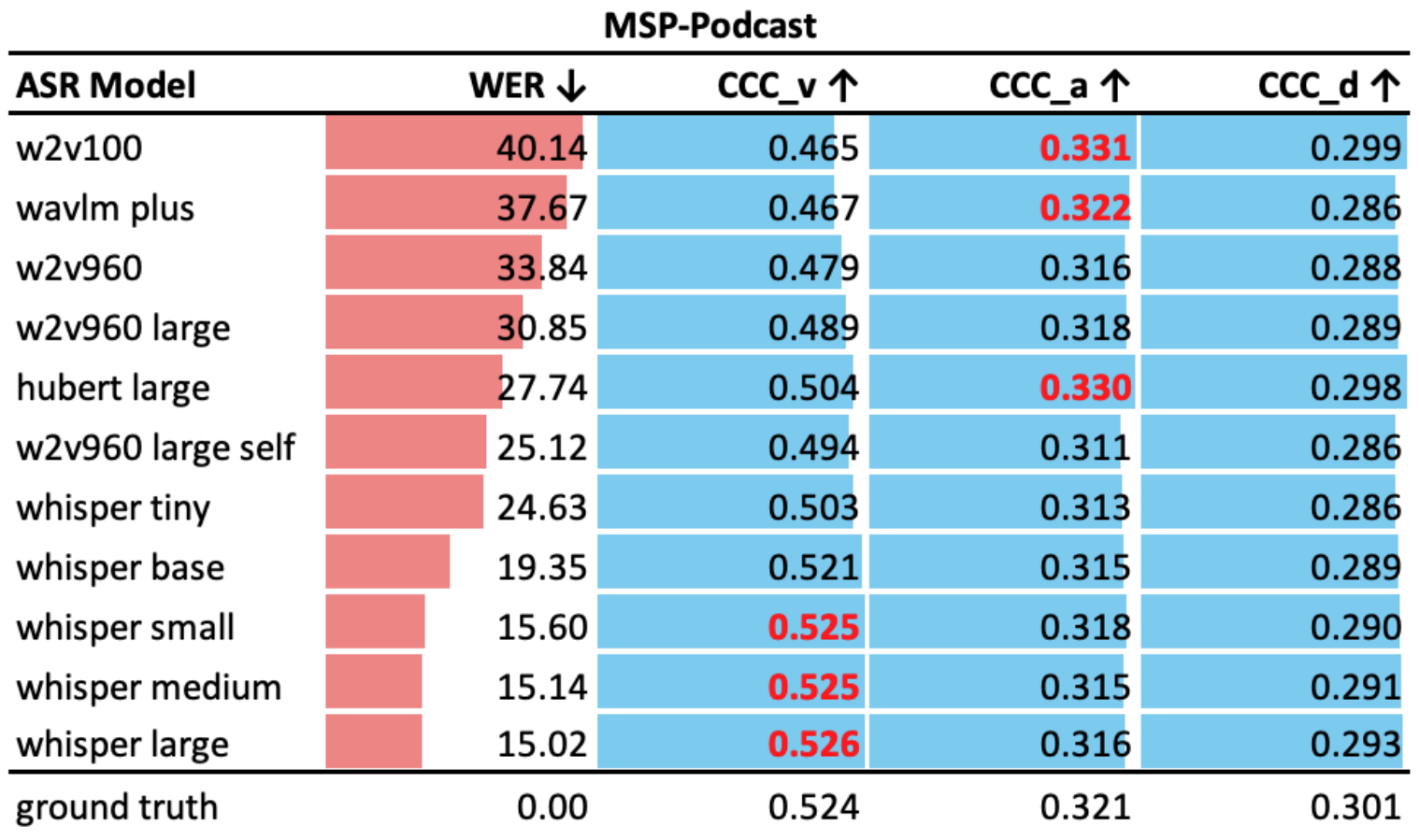}
    \label{tab:wer_ser}
    \vspace{-7pt}
\end{table}

\begin{table*}[ht!]
  \centering
  \caption{Benchmarking SER performance based on ASR transcripts with fusion techniques. Maximum diff: maximum difference between the performance of ground truth and that of ASR transcripts. \textbf{\textcolor{red}{Red bold}}: better performance than the ground truth.}
  \includegraphics[width=0.75\textwidth]{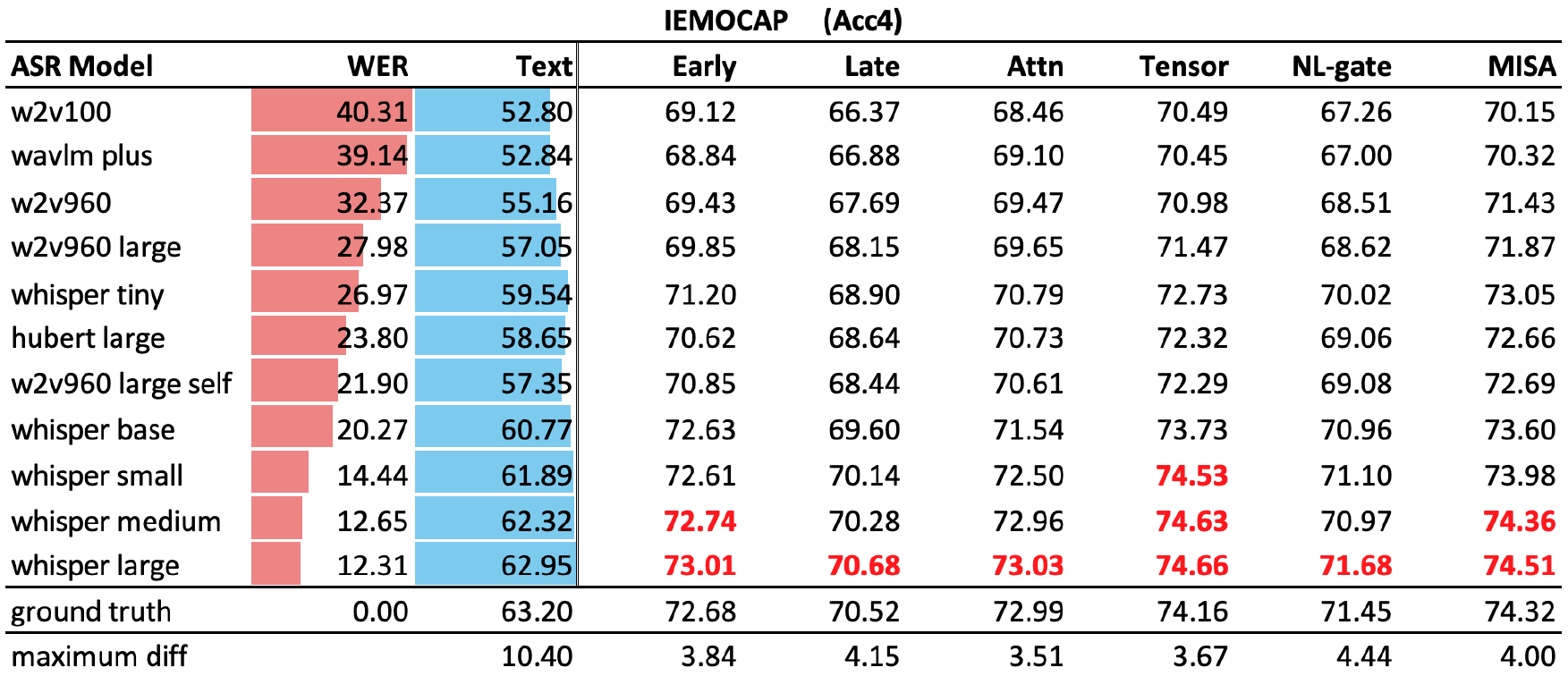}\vspace{3pt}
  \includegraphics[width=0.75\textwidth]{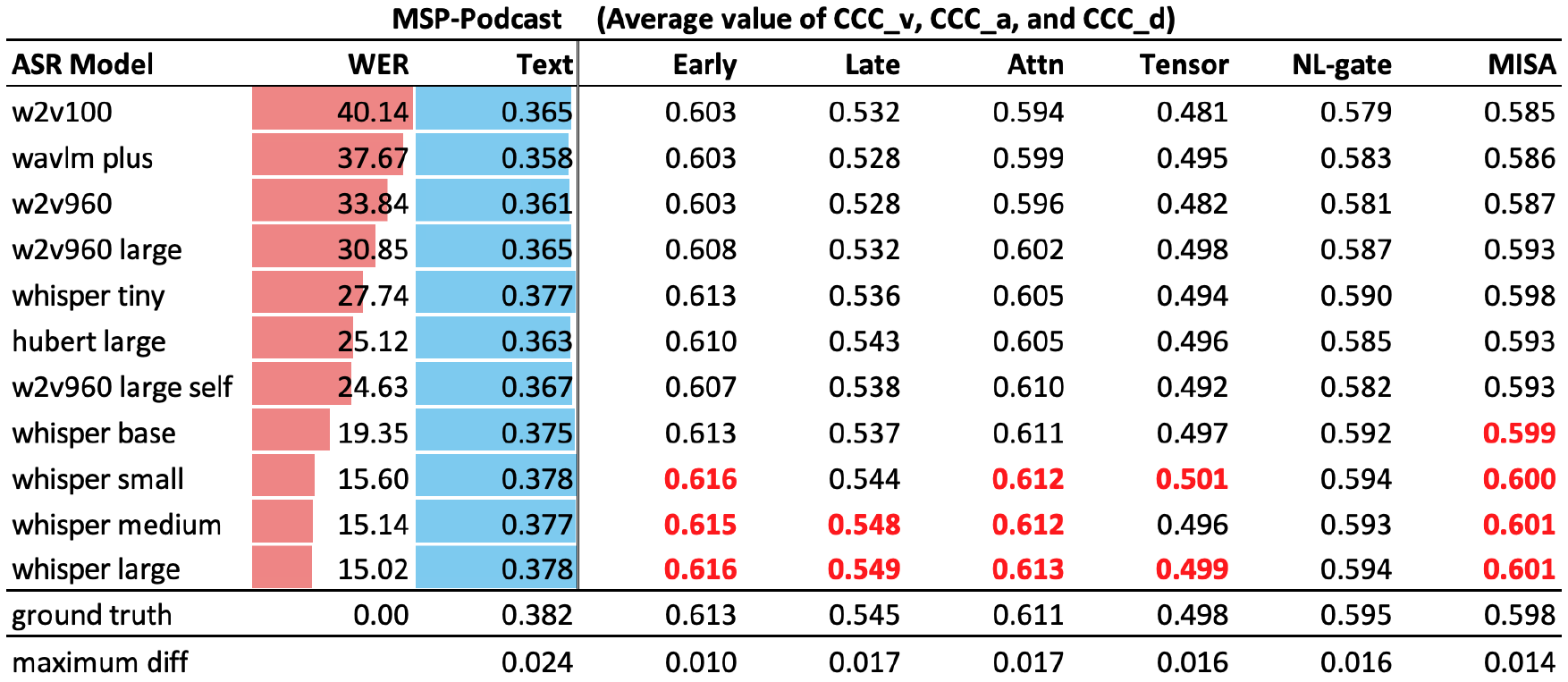}
  \label{tab:fusion}
  \vspace{-4pt}
\end{table*}

\textbf{2) SER is robust to relatively low WER, and in some cases, it is even better with ASR errors.} From IEMOCAP, it is observed that a WER of approximately 12\% has minimal impact on SER performance compared to ground-truth transcripts. Moreover, in CMU-MOSI, ASR errors can potentially enhance SER: transcripts generated by specific \textit{Whisper} models outperform those based on ground truth. Prior research has shown that sentiment analysis remains resilient to ASR errors \cite{tokuhisa2008emotion}, as positive or negative sentiments can encompass varied emotional states. Additionally, we find that in certain cases (e.g., relatively low WER), ASR errors do not diminish SER performance. Recent research on speech-based dementia detection indicates that ASR errors can offer valuable insights for dementia classification \cite{li2024useful}, possibly due to specific types of errors (e.g., repetitions, disfluencies) that may reveal indicators of dementia. However, we do not believe this applies similarly to SER. The phenomenon observed in our study may align with our previous finding: \textit{certain words being misrecognized as words that positively contribute to the ground-truth emotions.}

\textbf{3) Different metrics have different sensitivities to WER.} On CMU-MOSI, Acc7 shows a distinct pattern compared to Acc2 and MAE. Acc2 and MAE demonstrate consistent and smooth variations, whereas Acc7 appears random and lacks a discernible pattern. Interestingly, in nearly half of the transcripts (all from \textit{Whisper}), Acc7 outperforms the ground truth. This discrepancy may stem from the mismatch between the regression model used during training (since MOSI is labeled with continuous values) and the classification metric applied to Acc7, where predicted and ground-truth values are grounded. This mismatch could potentially blur accuracy and render SER insensitive to WER.

\textbf{4) Different emotion dimensions exhibit varying degrees of robustness to ASR errors.} From MSP-Podcast, it is evident that valence, arousal, and dominance exhibit distinct patterns. Firstly, the CCC of valence mirrors the pattern observed in Acc4 for IEMOCAP and Acc2 for CMU-MOSI, suggesting that valence shares similarities with categorical emotion in terms of robustness to ASR errors. Given that valence is conceptually similar to sentiment (indicating positivity or negativity), this alignment is plausible. 
Secondly, arousal and dominance do not exhibit a clear correlation with WERs. Since we jointly predict valence, arousal, and dominance, there may be shared information among these dimensions, resulting in minimal changes in arousal and dominance. However, our observations are consistent with the conventional understanding that valence is more influenced by textual content, whereas arousal and dominance are more influenced by audio cues \cite{schuller2009acoustic, wagner2023dawn, li2019expressing}.
Indeed, when we replaced text features with middle-layer audio features from \textit{W2V960-base} in the SER model, we obtained CCC\_v, CCC\_a, and CCC\_d values of 0.531, 0.635, and 0.558, respectively, further validating our observations.

\subsection{Benchmarking SER with WER and Fusion}
Next, we integrate audio features to investigate the robustness of existing fusion techniques in handling ASR errors in real-world scenarios. Since audio is not the primary focus of this study, we simply use the middle-layer audio features from \textit{W2V960-base}. For fusion techniques, we employ the following approaches: \\
\textbullet\ \textit{\textbf{Early fusion}}: text and audio features are concatenated at the embedding level. \\
\textbullet\ \textit{\textbf{Late fusion}}: Text and audio features are learned independently by their respective models, and the final decision is determined based on their outputs. \\
\textbullet\ \textit{\textbf{Cross-attention fusion}}: text and audio features are attended to each other via attention mechanism and then concatenated. \\
\textbullet\ \textit{\textbf{Tensor fusion}} \cite{zadeh2017tensor}: unimodal information and bimodal interactions are learned explicitly and then aggregated. \\
\textbullet\ \textit{\textbf{Non-local gate-based (NL-gate) Fusion}} \cite{wang2018non}: NL-gate is incorporated with multiple directions and at multiple positions with query-key-value within the attention mechanism. \\
\textbullet\ \textit{\textbf{Modality-invariant and -specific fusion (MISA)}} \cite{hazarika2020misa}: combining both modality-invariant and modality-specific features.

For fairness, we keep the backbone SER model used in the previous section unchanged, modifying only the input dimension of the first dense layer to match the output dimension of the hidden states from each fusion model. Additionally, we modify \textit{Tensor Fusion} and \textit{MISA} to receive bimodal inputs, as they were originally designed for trimodal inputs. Due to limited space, we present results only for IEMOCAP and MSP-Podcast for brevity, as these two corpora are sufficient to cover both categorical and dimensional emotions. Regarding the patterns observed in CMU-MOSI (omitted here), integrating audio did not significantly improve performance, as text alone already yields powerful results. Moreover, the audio in CMU-MOSI does not strongly convey emotion, leading to incongruity and ambiguity, which may pose challenges for SER in certain cases. From Table~\ref{tab:fusion}, it can be seen that:

\textbf{1) Fusing audio features largely mitigates the negative impact of increasing WER.} The decrease in Acc4 based on WER reaches 10\% without fusion on IEMOCAP, but only 4\% with fusion. Moreover, with most fusion techniques, the SER performance is even better on transcripts with relatively low WER than on ground truth. These findings confirm the benefits of bimodal SER in real applications, which can be achieved by using a pre-trained ASR model as a base for jointly generating transcripts and audio features \cite{li2022fusing, cai2021speech}.

\textbf{2) There is no optimal WER-robust fusion approach.} Due to different amounts of encoded information, the performances of fusion techniques are not directly comparable. For instance, \textit{Tensor Fusion} and \textit{MISA} encode more information by incorporating both unimodal and bimodal features, or both modality-invariant and -specific features. However, their effectiveness varies across different corpora. For example, they perform well on IEMOCAP (categorical) but poorly on MSP-Podcast (dimensional), indicating significant inconsistency. Thus, it is difficult to identify a single best-performing technique that is universally effective in addressing ASR errors. Nevertheless, the maximum difference between the performance of ground truth and that of ASR transcripts is still indicative of the robustness of the fusion techniques. We observe that although \textit{cross-attention fusion} may not produce the best performance, it yields a relatively small maximum difference, suggesting that it is less affected by WER. This observation is reasonable since cross-attention captures the relatedness between bimodal inputs, which changes dynamically once the words in text have changed, ensuring that the relatedness is least affected by ASR errors.

Since our experimental design impacts the results to some extent (e.g., removing punctuation, converting letters to lowercase, excluding samples without corresponding text transcription, modifying existing fusion models), it could inevitably lead to variations and exceptions in some SER performances. However, our overall findings and conclusions remain valid, especially as some of them are consistent with existing findings \cite{li2022fusing,he2024mf,feng2024foundation}.

\section{ASR Error-Robust Framework}
\begin{figure*}[ht!]
  \centering
  \includegraphics[width=0.92\textwidth]{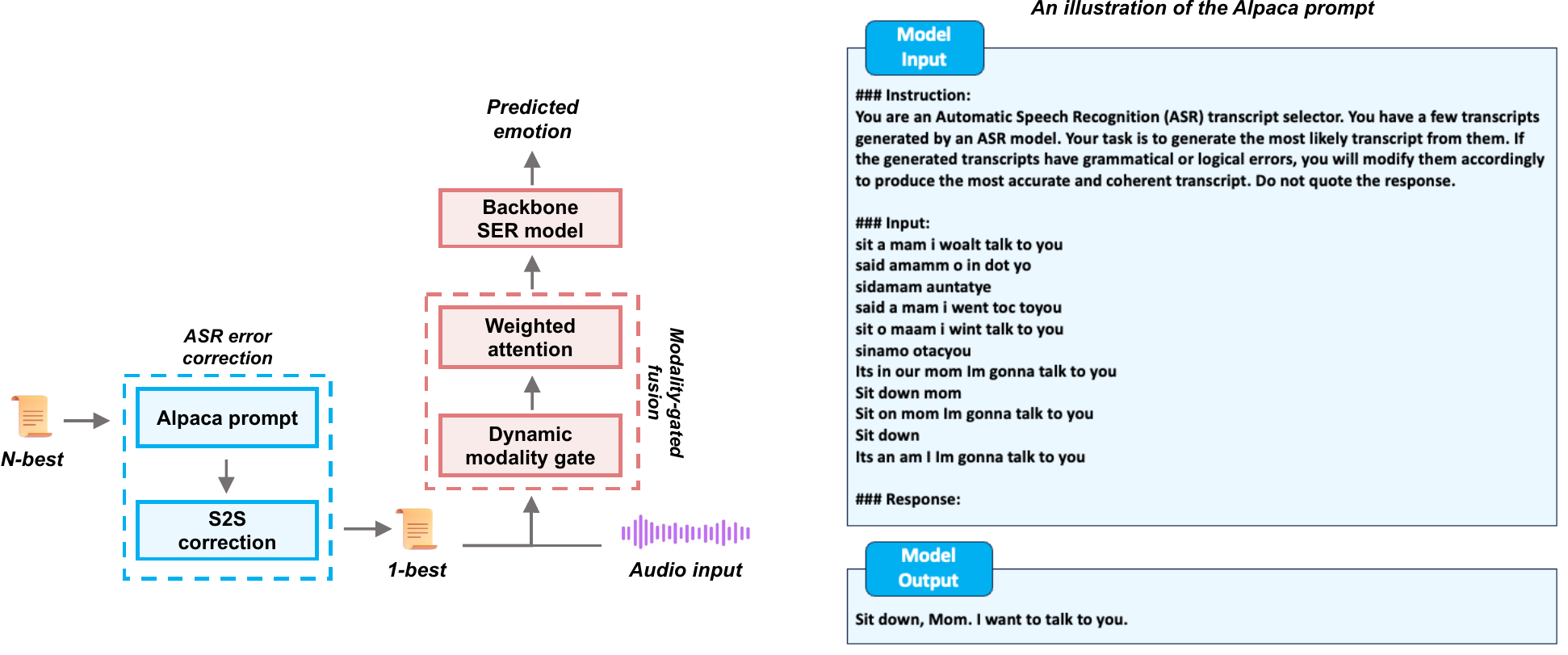}
  \caption{Proposed ASR error-robust framework. \textcolor{cyan}{\textbf{Blue}}: ASR error correction; frozen. \textcolor{pink}{\textbf{Pink}}: Modality-gated fusion; trainable.}
  \label{fig:proposal}
\end{figure*}

To address both the WER and fusion issues, we propose an ASR error-robust framework involving ASR error correction and modality-gated fusion, as illustrated in Fig.~\ref{fig:proposal} (left side).

\subsection{ASR Error Correction}
We combine the transcripts from all used ASR models, forming N-best (N=11) hypotheses. Compared to using raw N-best hypotheses directly from a single ASR model, which offer limited diversity, such a combination offers a more diverse range of transcription results. Furthermore, this approach facilitates the comparison of SER performance with each ASR model, as depicted in Table~\ref{tab:fusion}.

Then, the ASR error correction process consists of a two-stage process. First, we utilize the Alpaca prompt \cite{taori2023alpaca}, which guides a Large Language Model (LLM) to generate the most likely transcript (i.e., 1-best hypothesis) from the N-best hypotheses and correct any grammatical or logical errors. An illustration of the Alpaca prompt is depicted on the right of Fig.~\ref{fig:proposal}. We employ InstructGPT, which has demonstrated effectiveness in LLM-based ASR error correction \cite{radhakrishnan2023whispering}.

In the second stage, we further refine the 1-best transcripts by utilizing a Sequence-to-Sequence (S2S) ASR error correction model \cite{li2024crossmodal}, which was pre-trained on the \textit{Whisper-tiny} output $y'$ (i.e., 1-best hypothesis) of Common Voice \cite{ardila2019common} to recover the corrected sequence $y$ (i.e., gold annotation), with an optimizable parameter $\theta$:
\begin{align}
y=\arg\!\max_{y}P(y| y',\theta)
\end{align}

\subsection{Modality-Gated Fusion}
Based on the quality-enhanced transcripts, we adapt a trimodal incongruity-aware fusion method \cite{wang2023crossattention} to our bimodal inputs, considering potential distortion of original emotion in the spoken content due to ASR errors, which may introduce incongruence between emotions in audio and text.

Specifically, each modality is initially assigned an equal trainable weight. These weights dynamically adjust during training based on the respective contribution of each task, with larger contributions resulting in higher values. The sum of the weights always equals 1, and they are updated in every training batch to ensure dynamic adaptation to any input batch size. Subsequently, the weights are multiplied by corresponding modality features to perform weighted cross-attention fusion.
The pseudo-code is shown in Algorithm~\ref{alg}, where $H$ represents the hidden representation as input to the backbone SER model. $CrossAttn$ and $SelfAttn$ denote multihead attention, the former with the first argument as query and the second as key and value, and the latter with the argument as query, key, and value. The head number is set to eight.

\begin{algorithm}[!ht]
    \small
    \SetKwFunction{isOddNumber}{isOddNumber}
    \SetKwInOut{KwIn}{Input}

    \KwIn{
    Audio modality $A$; Text modality $T$;
    Dynamic modality gating weight $W^{1}$ initialized with \texttt{nn.Param(torch.tensor([1, 1]))};
    Feature concatenation weight $W^{2}$ initialized with \texttt{nn.Param(torch.tensor([1, 1, 1]))}}  
   
    \For{model input}
    {
        
        Sample $A$, $T$;
        
        $W^{1} \leftarrow \texttt{Softmax(}W^{1}\texttt{)}$;

        $W^{2} \leftarrow \texttt{Softmax(}W^{2}\texttt{)}$;

        $W^{1}_{A} \leftarrow W^{1}[0]$, $W^{1}_{T} \leftarrow W^{1}[1]$;

        $W^{2}_{A} \leftarrow W^{2}[0]$, $W^{2}_{T} \leftarrow W^{2}[1]$, $W^{2}_{AT} \leftarrow W^{2}[2]$,
                
        \eIf{$\texttt{argmax(}W'\texttt{)}$ == 0}
        {
            $A' = W^{1}_{A} * A$;

            $T' = W^{1}_{T} * CrossAttn(A, T)$;
        }{        
            $A' = W^{1}_{A} * CrossAttn(T, A)$;

            $T' = W^{1}_{T} * T$;
        }

        $H = SelfAttn(Concat\left[A'; T'\right])$;
                
        $H' = Concat\left[W^{2}_{A} * A, W^{2}_{T} * T, W^{2}_{AT} * H\right]$;
        
    }
    \caption{Modality-gated fusion.}
    \label{alg}
\end{algorithm}

Finally, we apply the same backbone SER model, keeping all other settings unchanged (e.g., the text and audio encoders) as in Section 4. We believe that this combination of dynamic modality gate and weighted attention can effectively mitigate the negative impact of ASR errors by consistently focusing on the most salient parts in both inputs for SER.

\subsection{Results and Discussion}

The results are presented in Table~\ref{tab:result}. We include only the MAE on CMU-MOSI and the average CCC on MSP-Podcast for brevity. Our approach further reduces WER and improves SER performance, demonstrating its effectiveness. While it does not surpass the best transcript in terms of MAE on CMU-MOSI, the values are very close. Moreover, in line with the findings in \cite{he2024mf}, our results confirm that employing proper ASR error-robust approaches can exceed SER performance based on ground-truth text using ASR transcripts. This phenomenon likely occurs because certain misrecognized words are interpreted as contributing positively to the ground-truth emotions, rather than ASR errors being preferable in SER.

\begin{table}[ht!]
\centering
\caption{Performance comparison. $\uparrow$: higher the better. $\downarrow$: lower the better.}
\scalebox{0.8}{
\begin{tabular}{lrr|rr|rr}
\multicolumn{1}{c}{\textbf{}} & \multicolumn{2}{c|}{\textbf{IEMOCAP}} & \multicolumn{2}{c|}{\textbf{CMU-MOSI}} & \multicolumn{2}{c}{\textbf{MSP-Podcast}} \\ \hline
 & \textbf{WER$\downarrow$} & \textbf{Acc4$\uparrow$} & \textbf{WER$\downarrow$} & \textbf{MAE$\downarrow$} & \textbf{WER$\downarrow$} & \textbf{CCC$\uparrow$} \\ \hline
best trans & 12.31 & 74.66 & 18.69 & 0.8558 & 15.02 & 0.616 \\
ground truth & 0.00 & 74.32 & 0.00 & 0.8902 & 0.00 & 0.613 \\ \hline
ours & \textbf{10.12} & \textbf{76.66} & \textbf{17.00} & \textbf{0.8557} & \textbf{12.85} & \textbf{0.618}
\end{tabular}
}
\label{tab:result}
\end{table}

\section{Conclusion}
In this study, we conducted a benchmark of SER performance using ASR transcripts with varying WERs and explored mainstream fusion techniques to assess the impact of ASR performance on SER. Our findings revealed several novel insights: 1) SER can tolerate relatively low WERs, especially in real-life speech scenarios. 2) Bimodal SER with transcripts containing approximately 10\% errors may not perform worse than those with ground-truth text, particularly with powerful Whisper models. However, further analysis is necessary to understand the nature and locations of ASR errors, as well as the mechanisms underlying fusion techniques. Moreover, we proposed an ASR error-robust framework that integrates ASR error correction and modality-gated fusion. This framework demonstrated superior performance compared to baseline models in our experimental results and holds potential for various tasks where ASR transcription serves as the text source. In the near future, we plan to launch a challenge aimed at addressing ASR errors for robust multimodal SER.

\bibliographystyle{IEEEbib}
\bibliography{refs}

\end{document}